\documentclass[aps,
twocolumn,
superscriptaddress,showpacs]{revtex4}
\usepackage{epsf}
\usepackage{graphicx}
\usepackage{amsmath}
\begin{document}
\title{Thermodynamics of Peptide Aggregation Processes. \\An Analysis
from Perspectives of Three Statistical Ensembles}
\author{Christoph Junghans}
\email[E-mail: ]{junghans@mpip-mainz.mpg.de}
\affiliation{Institut f\"ur Theoretische Physik and Centre for Theoretical Sciences (NTZ), 
Universit\"at Leipzig, Postfach 100\,920, D-04009 Leipzig, Germany}
\affiliation{Max-Planck-Institut f\"ur Polymerforschung, Ackermannweg 10, D-55128 Mainz, Germany}
\author{Michael Bachmann}
\email[E-mail: ]{Michael.Bachmann@itp.uni-leipzig.de}
\affiliation{Institut f\"ur Theoretische Physik and Centre for Theoretical Sciences (NTZ), 
Universit\"at Leipzig, Postfach 100\,920, D-04009 Leipzig, Germany}
\affiliation{Computational Biology \& Biological Physics Group, Department of Theoretical Physics,  
Lunds Universitet, S\"olvegatan 14A, SE-223\,62 Lund, Sweden}
\author{Wolfhard Janke}
\email[E-mail: ]{Wolfhard.Janke@itp.uni-leipzig.de}
\homepage[\\ Homepage: ]{http://www.physik.uni-leipzig.de/CQT.html}
\affiliation{Institut f\"ur Theoretische Physik and Centre for Theoretical Sciences (NTZ), 
Universit\"at Leipzig, Postfach 100\,920, D-04009 Leipzig, Germany}
\begin{abstract}
We employ a mesoscopic model for studying aggregation processes of protein-like hydrophobic-polar 
heteropolymers. By means of multicanonical Monte Carlo computer simulations, we find strong 
indications that peptide aggregation is a phase separation process, in which the
microcanonical entropy exhibits a convex intruder due to nonnegligible surface effects of the small
systems.
We analyze thermodynamic properties of the conformational transitions accompanying the aggregation
process from the multicanonical, canonical, and microcanonical perspective. It turns out that
the microcanonical description is particularly advantageous as it allows for unraveling details of
the phase-separation transition in the thermodynamic region, where the temperature is not a suitable 
external control parameter anymore.
\end{abstract}
\maketitle
\section{Introduction}
Beside receptor-ligand binding mechanisms, folding and aggregation of proteins
belong to the biologically most relevant molecular structure formation processes. While the
specific binding between receptors and ligands is not necessarily accompanied by global cooperative
structural changes, protein folding and oligomerization of peptides are typically accompanied 
by conformational transitions~\cite{gsponer1}. Proteins and their aggregates are comparatively small systems.
A typical protein consists of a sequence of some hundred amino acids and aggregates are often
formed by only a few peptides. A very prominent example is the extracellular aggregation of the 
A$\beta$ peptide, which is associated with Alzheimer's disease. Following the amyloid hypothesis, 
it is believed that these aggregates (which can also take fibrillar
forms) are neurotoxic, i.e., they are able to fuse into cell membranes of neurons and open calcium
ion channels. It is known that extracellular Ca$^{2+}$ ions intruding into a neuron can promote its
degeneration~\cite{lin1,quist1,lashuel1}.

Conformational transitions proteins experience
during structuring and aggregation are not phase transitions in the strict thermodynamic sense
and their statistical analysis is usually based on studies of signals exposed by energetic 
and structural fluctuations, as well as system-specific ``order'' parameters. In these studies, 
the temperature $T$ is considered as an adjustable,
external control parameter and, for the analysis of the pseudophase transitions, the peak structure 
of quantities such as the specific heat and the fluctuations of the gyration tensor components or 
``order'' parameter as functions of the temperature are investigated. The natural ensemble for 
this kind of analysis is the canonical ensemble, where the possible states of the 
system with energies $E$ are distributed according to the Boltzmann probability
$\exp(-E/k_BT)$, where $k_B$ is the Boltzmann constant. However, 
phase separation processes of small systems as, e.g., droplet condensation, are accompanied 
by surface effects at the interface between the 
pseudophases~\cite{gross1,gross2,thirring1,schmidt1,pichon1,lopez1,wj1,pleimling1,wales1,hilbert1,nussbaumer1}. 
This is reflected by the behavior
of the microcanonical entropy ${\cal S}(E)$, which exhibits a \textit{convex} monotony in the
transition region. Consequences are the backbending of the caloric temperature
$T(E)=(\partial {\cal S}/\partial E)^{-1}$, i.e., the \textit{decrease} of temperature 
with increasing system energy,
and the negativity of the microcanonical specific heat $C_V(E)=(\partial T(E)/\partial E)^{-1}=
-(\partial {\cal S}/\partial E)^2/(\partial^2 {\cal S}/\partial E^2)$. The physical reason is that the free energy 
balance in phase equilibrium requires the minimization of the interfacial surface and, therefore,
the loss of entropy~\cite{gross3}. 
A reduction of the entropy can, however, only be achieved by transferring energy 
into the system.  
Recently, we have shown that, employing a minimalistic heteropolymer model, the aggregation 
of two small peptides is such a phase separation process, where we observed the mentioned peculiar 
small-system effects~\cite{jbj1}. Here, we consider the aggregation process from  
the multicanonical, canonical, and microcanonical perspectives. 
Our results were obtained from multicanonical computer simulations of a mesoscopic hydrophobic-polar
heteropolymer model for aggregation, which is based on a simple off-lattice model, originally 
introduced to study tertiary folding of proteins from a coarse-grained point of view.

The paper is organized as follows. In Sect.~\ref{sec:modmet}, we define the aggregation model
employed in our computational study, where we primarily used multicanonical sampling. This method
is also briefly described here as well as the aggregation ``order'' parameter needed to 
discriminate the pseudophases. 
Section~\ref{sec:twochain} is devoted to the main part of the paper: The presentation of the results
for the aggregation of two small peptides obtained from multicanonical, canonical, and microcanonical
views. The comparison with results obtained for larger systems is performed in Sect.~\ref{sec:larger}.
The paper is concluded by a summary of the results in Sect.~\ref{sec:sum}. 
\section{Model and methods}
\label{sec:modmet}
For our aggregation study on mesoscopic scales, we employ a novel
model that is based on a known hydrophobic-polar single-chain approach, originally introduced
for heteropolymer chains in two dimensions~\cite{still1}. In this section, we define this model,
describe the simulation methods, and introduce a suitable order parameter that allows
for the discrimination of the macrostates or ``pseudophases'' the multiple-chain system can reside 
in.
\subsection{Mesoscopic hydrophobic-polar aggregation model}
\label{sec:model}
For our aggregation study of protein-like heteropolymers, we assume that the tertiary folding
process of the individual chains is governed by hydrophobic-core formation in an aqueous 
environment. A comparatively simple but powerful model is the AB model~\cite{still1} where only two
types of amino acids are considered: hydrophobic residues (A) which avoid contact with the polar
environment and polar residues (B) being favorably attracted by the solvent. The model is a 
C$^\alpha$ type model in that each residue is represented by only a single interaction site (the
``C$^\alpha$ atom''). Thus, the natural dihedral torsional degrees of freedom of realistic 
protein backbones are replaced by virtual bond and torsion angles between consecutive
interaction sites. The large torsional barrier
of the peptide bond between neighboring amino acids is in the AB model effectively taken into 
account by introducing a bending energy. Nonbonded residues experience weak pairwise
long-range attraction ($AA$ and $BB$ pairs)
or repulsion ($AB$ pairs), respectively. Although this coarse-grained picture is obviously not 
capable to reproduce microscopic properties of specific realistic proteins, it qualitatively
exhibits, however, sequence-dependent features known from nature, as, for example, tertiary folding
pathways known from two-state folding, folding through intermediates, and metastability~\cite{ssbj1}.

For our systems of more than one chain, we further assume that the interaction strengths
between nonbonded residues is independent of the individual properties of the chains the
residues belong to. Therefore, we use the same parameter sets as in the AB model for the 
pairwise interactions between residues of different chains. 
Our aggregation model reads~\cite{jbj1}
\begin{equation}
\label{eq:aggmod}
E=\sum\limits_{\mu} E_{\rm AB}^{(\mu)}+\sum\limits_{\mu<\nu} 
\sum_{i_\mu,j_\nu}\Phi(r_{i_\mu j_\nu};\sigma_{i_\mu},\sigma_{j_\nu}),
\end{equation}
where $\mu,\nu$ label the $M$ polymers interacting with each other, and 
$i_\mu,j_\nu$ index the $N_{\mu,\nu}$ monomers of the respective $\mu$th and $\nu$th polymer.
The intrinsic single-chain energy is given by
\begin{equation}
\label{eq:abmod}
E_{\rm AB}^{(\mu)}=\frac{1}{4}\sum\limits_{i_\mu}(1-\cos \vartheta_{i_\mu})+%
\!\!\sum\limits_{j_\mu>i_\mu+1}\Phi(r_{i_\mu j_\mu};\sigma_{i_\mu},\sigma_{j_\mu}),
\end{equation}
with $0\le \vartheta_{i_\mu}\le \pi$ denoting the bending angle between monomers 
$i_\mu$, $i_\mu+1$, and $i_\mu+2$.
The nonbonded inter-residue pair potential 
\begin{equation}
\label{eq:phi}
\Phi(r_{i_\mu j_\nu};\sigma_{i_\mu},\sigma_{j_\nu})=
4\left[r_{i_\mu j_\nu}^{-12}-C(\sigma_{i_\mu},\sigma_{j_\nu})r_{i_\mu j_\nu}^{-6}\right]
\end{equation}
depends on the distance $r_{i_\mu j_\nu}$ between the residues, and on their type,
$\sigma_{i_\mu}=A,B$. The long-range behavior is attractive for 
like pairs of residues [$C(A,A)=1$, $C(B,B)=0.5$] and repulsive otherwise [$C(A,B)=C(B,A)=-0.5$]. 
The lengths of all virtual peptide bonds are set to unity. 

In this paper, we report on results obtained from statistical mechanics studies of the 
aggregation processes of short polymers. Our primary interest is devoted to the heteropolymer 
with the sequence S1: $AB_2AB_2ABAB_2AB$ which is a Fibonacci sequence~\cite{still1},
whose single-chain properties are already known~\cite{baj1}. Throughout the paper, we 
are going to study the thermodynamics 
of systems with up to 4 chains of this sequence over the whole energy and temperature 
regime. 
\subsection{Simulation methods}
\label{sec:sim}
We have used generalized-ensemble Markovian Monte Carlo algorithms to sample the conformational 
space of the systems studied. The powerful error-weighted multicanonical method~\cite{muca1,muca2,muca3} 
proved to be particularly useful as it makes it possible to scan the whole phase space 
with very high accuracy~\cite{baj1}. The principle idea is to deform the Boltzmann energy 
distribution 
\begin{equation}
\label{eq:can}
p_{\rm can}(E;T)\propto g(E)\exp(-E/k_BT), 
\end{equation}
where $g(E)$ is the density of states with energy
$E$ and $k_BT$ is the thermal energy at temperature $T$, in such a way that the notoriously difficult
sampling of the tails is increased and -- particularly useful -- the sampling rate of the entropically
strongly suppressed lowest-energy conformations is improved. In order to achieve this, the
canonical Boltzmann distribution is modified by the multicanonical weight $W_{\rm muca}(E;T)$
which, in the ideal case, flattens the energy distribution:
\begin{equation}
\label{eq:muca}
p_\text{muca}(E)=W_\text{muca}(E;T)p_{\rm can}(E;T)=\text{const}_{E;T}.
\end{equation}
As the canonical distribution is, of course, not known 
in the beginning and 
$W_\text{muca}(E;T)\sim p_{\rm can}^{-1}(E;T)$, the multicanonical weights 
have to be determined 
recursively, which can be done in an efficient way~\cite{muca3,celik1}. Recalling that the 
simulation temperature $T$ does not
possess any meaning in the multicanonical ensemble as, according to Eq.~(\ref{eq:muca}), the 
energy distribution is always constant, independently of temperature. 
Actually, it is convenient to set it to infinity in which
case $\lim_{T\to\infty}p_{\rm can}(E;T)\sim g(E)$ and thus 
$\lim_{T\to\infty}W_{\rm muca}(E;T)\sim g^{-1}(E)$. The latter
expression is sometimes parametrized as $W_\text{muca}(E)\sim\exp[-\beta(E)E+\alpha(E)]$,
where, for a suitable choice of $\alpha(E)$, $\beta(E)$ can be identified with the 
microcanonical temperature~\cite{celik1}.

%

In our simulations, conformational changes
of the individual chains included spherical updates~\cite{baj1} and semilocal crankshaft moves, i.e., 
rotations around the axis between the $n$th and $(n+2)$th residue. A typical multicanonical run
contained of the order of $10^{10}$ single updates. The polymer chains were embedded into a cubic box
with edge lengths $L$ and periodic boundary conditions were used. 
In our simulations, the edge lengths of the simulation box were chosen to be $L=40$ which is sufficient to
reduce undesired finite-size effects.

For cross-checks we have also performed replica-exchange (parallel tempering) simulations~\cite{pt1,pt2}.
Verifying lowest-energy conformations found in the multicanonical simulations, we have also performed
optimization runs using the energy-landscape paving (ELP) method~\cite{elp}. 
\subsection{``Order'' parameter of aggregation and fluctuations}
\label{sec:aggpar}
In order to distinguish between the fragmented and the aggregated regime, we introduce 
the ``order'' parameter
\begin{equation}
\Gamma^2=\frac{1}{2M^2}\sum_{\mu,\nu=1}^{M} {\bf d}_{\rm per}^2\left({\bf r}_{{\rm COM},\mu},
{\bf r}_{{\rm COM},\nu}\right),
\end{equation}
where the summations are taken over the minimum distances 
${\bf d}_{\rm per}=\left(d_{\rm per}^{(1)},d_{\rm per}^{(2)},d_{\rm per}^{(3)}\right)$
of the respective  
centers of mass of the chains (or their periodic continuations).  
The center of mass of the $\mu$th chain in a box with periodic boundary conditions
is defined as ${\bf r}_{\text{COM},\mu}=\sum_{i_\mu=1}^{N_\mu} \left[
{\bf d}_{\rm per}\left({\bf r}_{i_\mu},{\bf r}_{1_\mu}\right)+{\bf r}_{1_\mu}\right]/N_\mu$, where
${\bf r}_{1_\mu}$ is the coordinate vector of the first monomer and serves as a reference coordinate 
in a local coordinate system. 

Our aggregation parameter is to be considered 
as a qualitative measure; roughly, fragmentation corresponds to large values of $\Gamma$, aggregation
requires the centers of masses to be in close distance in which case $\Gamma$ is comparatively small.
Despite its qualitative nature, it turns out to be a surprisingly manifest indicator for the 
aggregation transition and allows even a clear discrimination of different aggregation pathways, as
will be seen later on.

According to the Boltzmann distribution~(\ref{eq:can}), we define canonical expectation values 
of any observable $O$ by
\begin{eqnarray}
&&\hspace*{-8mm}\langle O\rangle(T)=\nonumber\\
&&\hspace*{-3mm}\frac{1}{Z_{\rm can}(T)}
\prod\limits_{\mu=1}^M\left[\int {\cal D}{\bf X}_\mu\right]O(\{{\bf X}_\mu\})
e^{-E(\{{\bf X}_\mu\})/k_BT},
\end{eqnarray}
where the canonical partition function $Z_{\rm can}$ is given by
\begin{equation}
Z_{\rm can}(T)=\prod\limits_{\mu=1}^M\left[\int {\cal D}{\bf X}_\mu\right]e^{-E(\{{\bf X}_\mu\})/k_BT}.
\end{equation}
Formally, the integrations are performed over all possible conformations ${\bf X}_\mu$ of the $M$
chains.

Similarly to the specific heat per monomer
$c_V(T)=d\langle E\rangle/N_{\rm tot}dT=(\langle E^2\rangle-\langle E\rangle^2)/N_{\rm tot}k_BT^2$
(with $N_{\rm tot}=\sum_{\mu=1}^M N_\mu$) 
which expresses the thermal fluctuations of energy, the temperature derivative of $\langle\Gamma\rangle$ per monomer,
$d\langle\Gamma\rangle/N_{\rm tot}dT=(\langle \Gamma E\rangle-\langle\Gamma\rangle\langle E\rangle)/N_{\rm tot}k_BT^2$, is 
a useful indicator for cooperative behavior of the multiple-chain system. 
Since the system size
is small -- the number of monomers $N_{\rm tot}$ as well as the number of chains $M$ -- aggregation transitions,
if any, are expected to be signalized by the peak structure of the fluctuating quantities as functions
of the temperature. This requires the temperature to be a unique external control parameter which
is a natural choice in the canonical statistical ensemble. Furthermore, this is a typically easily
adjustable and, therefore, convenient parameter in experiments. As we have stressed recently~\cite{jbj1}, 
however, aggregation is a phase separation process and, since the system is small, there is no uniform
mapping between temperature and energy. For this reason, the total system energy is the more appropriate
external parameter. Thus, the microcanonical interpretation will turn out to be the more favorable 
description, at least in the transition 
region. We will discuss this in detail in the following section.
\section{Statistics of the two-chain heteropolymer system in three ensembles}
\label{sec:twochain}
For the qualitative description of the aggregation and the accompanied conformational cooperativity
within the whole system, it is sufficient to consider a very small system which is 
computationally reliably tractable and nonetheless yields precise results for all energies and
temperatures. Our heteropolymer
system consists of two identical chains with the amino acid composition S1 and
will be denoted as 2$\times$S1.
In the following, we discuss the aggregation behavior
of this system from the multicanonical, the canonical, and the microcanonical point of view.
\subsection{Multicanonical results}
\label{sec:muca}
In a multicanonical simulation, the phase space is sampled in such a way that the energy distribution
gets as flat as possible. Thermodynamically, this means that the sampling of the phase space is
performed for all temperatures within a single simulation~\cite{muca1,muca2,muca3,celik1}. The desired information for the
thermodynamic behavior of the system at a certain temperature is then obtained by simply
reweighting the multicanonical into the respective canonical distribution, according to
Eq.~(\ref{eq:muca}). Since the multicanonical ensemble contains all thermodynamic informations, 
including the conformational transitions, it is quite useful to measure within the simulation 
the multicanonical histogram 
\begin{equation}
\label{eq:muhist}
h_{\rm muca}(E_0,\Gamma_0)=\sum\limits_{t_{\rm muca}}\delta_{E,E_0}\delta_{\Gamma,\Gamma_0},
\end{equation}
where $t_{\rm muca}$ labels the Monte Carlo ``time'' steps. More formally, this distribution 
can be expressed as  
a conformation-space integral
\begin{eqnarray}
\label{eq:muhist2}
&&\hspace*{-4mm}h_{\rm muca}(E_0,\Gamma_0)\propto\langle\delta(E-E_0)\delta(\Gamma-\Gamma_0)
\rangle_{\rm muca}\nonumber\\
&&\hspace*{-4mm}=\frac{1}{Z_{\rm muca}}\prod_{\mu=1}^M\left[\int {\cal D}{\bf X}_\mu\right] 
\delta(E(\{{\bf X}_\mu\}-E_0)\delta(\Gamma(\{{\bf X}_\mu\}-\Gamma_0)\nonumber\\
&&\times e^{-{\cal H}_{\rm muca}(E(\{{\bf X}_\mu\}))/k_BT}
\propto e^{-{\cal F}_{\rm muca}(E_0,\Gamma_0)/k_BT}
\end{eqnarray}
with the multicanonical energy ${\cal H}_{\rm muca}(E)=E-k_BT\ln\,W_{\rm muca}(E;T)$ 
which is independent 
of temperature.

The multicanonical partition function is also trivially a constant in temperature,
\begin{equation}
Z_{\rm muca}=\prod_{\mu=1}^M\left[\int {\cal D}{\bf X}_\mu\right]
e^{-{\cal H}_{\rm muca}(E(\{{\bf X}_\mu\}))/k_BT}=\text{const}_T.
\end{equation}
It is obvious that integrating $h_{\rm muca}(E,\Gamma)$ over $\Gamma$ recovers the uniform 
multicanonical energy distribution:
\begin{equation}
\int\limits_0^\infty d\Gamma\, h_{\rm muca}(E,\Gamma) \sim p_{\rm muca}(E).
\end{equation}
The canonical distribution of energy and $\Gamma$ parameter at temperature $T$ 
can be retained, similar to inverting Eq.~(\ref{eq:muca}), by performing the simple reweighting
\begin{equation}
h_{\rm can}(E,\Gamma;T) = h_{\rm muca}(E,\Gamma)W^{-1}_{\rm muca}(E;T),
\end{equation}
which is, due to the restriction to a certain temperature, less favorable to gain an overall 
impression of the phase behavior (i.e., the transition pathway) of
the system, compared to the multicanonical analog $h_{\rm muca}(E,\Gamma)$.

\begin{figure}
\centerline{\epsfxsize = 8.8cm \epsfbox{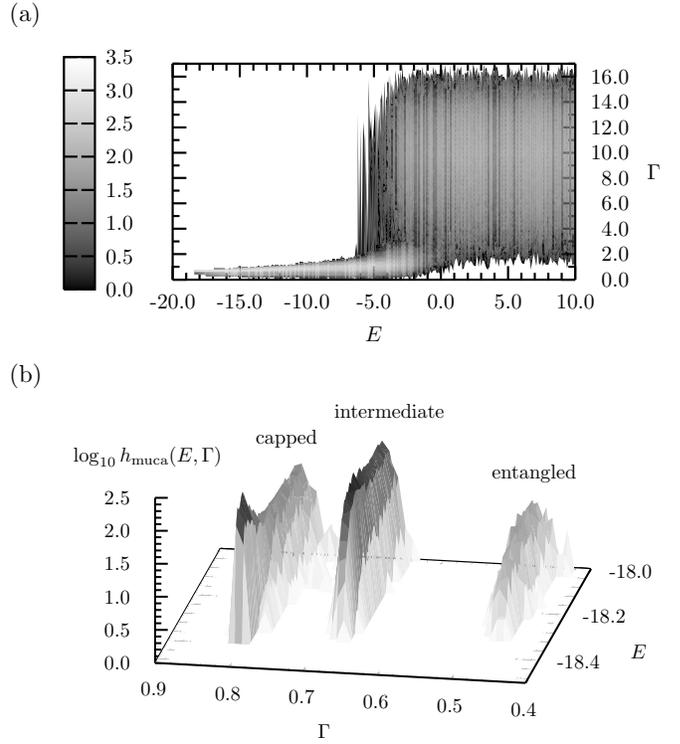}}
\caption{\label{fig:mucahist} (a) Multicanonical histogram $\log_{10}h_{\rm muca}$ as
a function of energy $E$ and aggregation parameter $\Gamma$, (b) section of $\log_{10}h_{\rm muca}$
in the low-energy tail.}
\end{figure}
In Fig.~\ref{fig:mucahist}(a), $h_{\rm muca}(E,\Gamma)$ is shown for the two-peptide system 2$\times$S1 as
a color-coded projection onto the $E$-$\Gamma$ plane, which is the direct output obtained 
in the multicanonical 
simulation. Qualitatively, we observe two separate main branches (which are ``channels'' in the corresponding
free-energy landscape), between which a noticeable
transition occurs. In the vicinity of the energy $E_{\rm sep}\approx -3.15$, both channels overlap, i.e., 
the associated macrostates coexist. Since $\Gamma$ is an effective measure for the spatial distance between
the two peptides, it is obvious that conformations with separated or fragmented peptides belong to 
the dominating channel in the regime of high energies and large $\Gamma$ values, whereas the aggregates are
accumulated in the narrow low-energy and small-$\Gamma$ channel. Thus, the main observation from the 
multicanonical, comprising point of view is that the aggregation transition is a phase-separation process
which, even for this small system, already appears in a surprisingly clear fashion. 

\begin{figure}
\centerline{\epsfxsize = 7.5cm \epsfbox{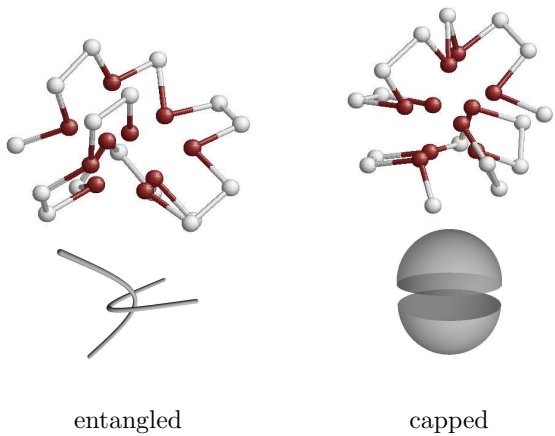}}
\caption{\label{fig:confs} Representatives and schematic characteristics of entangled and 
spherically-capped conformations dominating the lowest-energy branches in the multicanonical
histogram shown in Fig.~\ref{fig:mucahist}(b). Dark spheres correspond to hydrophobic ($A$), 
light ones to polar ($B$) residues.}
\end{figure}
The high precision
of the multicanonical method allows us even to reveal further details in the lowest-energy aggregation regime,
which is usually a notoriously difficult sampling problem. Figure~\ref{fig:mucahist}(b) shows that the tight
aggregation channel splits into three separate, almost degenerate subchannels at lowest energies. From the analysis
of the conformations in this region, we find that representative conformations with smallest $\Gamma$ values,
$\Gamma\approx 0.45$, are typically entangled, while those with $\Gamma\approx 0.8$ have a spherically-capped
shape. This is the subchannel connected to the lowest-energy states. Examples are shown in Fig.~\ref{fig:confs}. The also highly compact conformations belonging to the 
intermediate subphase do not exhibit such 
characteristic features and are rather globules without noticeable internal symmetries. In all cases, 
the aggregates contain a single compact core of hydrophobic residues. Thus, the aggregation is not a simple
docking process of two prefolded peptides, but a complex cooperative folding-binding process. This is 
a consequence of the energetically favored hydrophobic inter-residue contacts which, as the results show,
overcompensate the entropic steric constraints. The story is, however, even more interesting, as also 
nonnegligible surface effects come into play. After the following standard
canonical analysis, this will be discussed in more detail in the subsequent 
microcanonical interpretation of our results.
\subsection{\label{sec:can}Canonical perspective} 
Phase transitions are typically described in the canonical ensemble
with the temperature kept fixed. This is also natural from an 
experimentalist's point of view, since the temperature is a convenient external control 
parameter. The macrostates are weighted according to the Boltzmann distribution~(\ref{eq:can}).
A nice feature of the canonical ensemble is that the temperature dependence of 
fluctuations of thermodynamic quantities
is usually a very useful indicator for phase or pseudophase
transitions. This cooperative thermodynamic activity is typically signalized by peaks or,
in the thermodynamic limit (if it exists), by divergences of these fluctuations. Even for
small systems, peak temperatures can frequently be identified with transition temperatures.
Although in these cases peak temperatures typically depend on the fluctuating quantities 
considered, in most cases associated pseudophase transitions are doubtlessly manifest. 
In such cases, the
transition ranges over an extended temperature interval, as, e.g., in the folding process
of proteins or heteropolymers~\cite{bj1}. 

\begin{figure}
\centerline{\epsfxsize=8.8cm \epsfbox{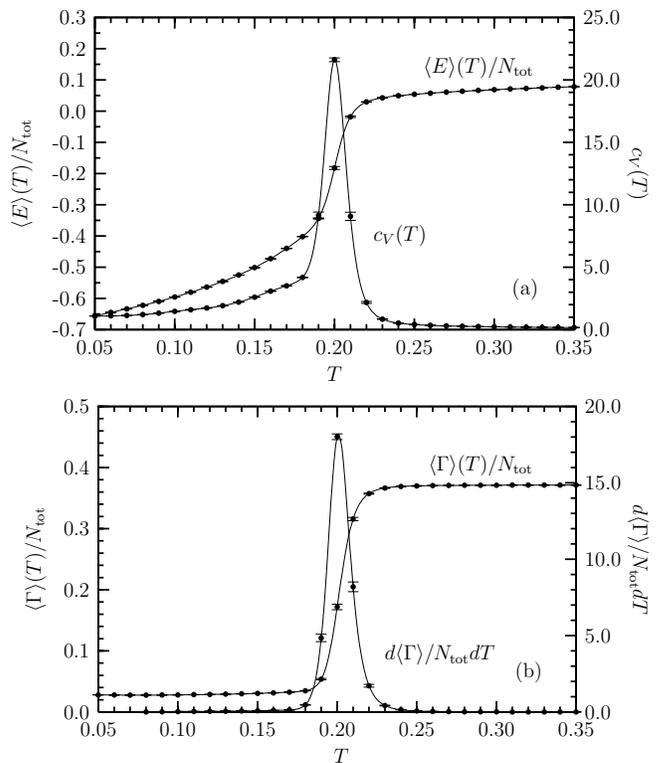}}
\caption{\label{fig:canon2x13.1} (a) Mean energy $\langle E\rangle/N_{\rm tot}$ and specific heat per monomer 
$c_V$, and (b) $\langle \Gamma\rangle/N_{\rm tot}$ and $d\langle\Gamma\rangle/N_{\rm tot}dT$ as functions
of the temperature.}
\end{figure}
In our aggregation study of the 2$\times$S1 system, however, we obtain from the canonical analysis
a surprisingly clear picture of the aggregation transition.
Figure~\ref{fig:canon2x13.1}(a) shows the canonical mean energy $\langle E\rangle$ and 
the specific heat per 
monomer $c_V$, plotted as functions of the temperature $T$. In Fig.~\ref{fig:canon2x13.1}(b), the 
temperature dependence of the mean aggregation order parameter $\langle \Gamma\rangle$ and 
the fluctuations of $\Gamma$ are shown. The aggregation transition is signalized by very sharp
peaks and from both figures we read off peak temperatures close to $T_{\rm agg}\approx 0.20$. The 
aggregation of the two peptides is a single-step process, in which the formation of the 
aggregate with a common compact hydrophobic core governs the folding behavior of the individual
chains. Folding and binding are not separate processes. 

\begin{figure}
\centerline{\epsfxsize=3.3cm \epsfbox{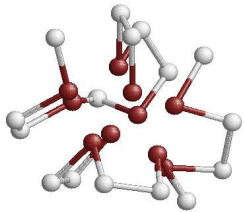}}
\caption{\label{fig:gsconf} The minimum-energy 2$\times$S1 complex with $E_{\rm min}\approx -18.4$ 
as found in our simulations is a capped aggregate.}
\end{figure}

The dominance of the inter-chain
binding interaction can also be seen by considering the lowest-energy conformation found 
in our simulations. The energy of this conformation, which is shown in Fig.~\ref{fig:gsconf},
is $E_{\rm min}\approx -18.4$ in our energy units. The peptide-peptide binding energy
[i.e., the second term in Eq.~(\ref{eq:aggmod})] is with 
$E_{AB,{\rm min}}^{(1,2)}\approx -11.4$ much stronger
than the intrinsic single-chain energies $E_{AB,{\rm min}}^{(1)}\approx -3.2$ and 
$E_{AB,{\rm min}}^{(2)}\approx -3.8$, respectively. The single-chain minimum energy
is with $E_{AB,{\rm min}}^{\rm single}\approx -5.0$~\cite{baj1} noticeably smaller.

The comparatively strong inter-chain interaction and the strength of the aggregation transition despite
the smallness of the system lead to the conclusion that surface effects are of essential
importance for the aggregation of the peptides. This is actually confirmed by a detailed 
microcanonical analysis which is performed in the next subsection.   
\subsection{\label{sec:micro} Microcanonical interpretation}
\begin{figure}
\centerline{\epsfxsize=8.8cm \epsfbox{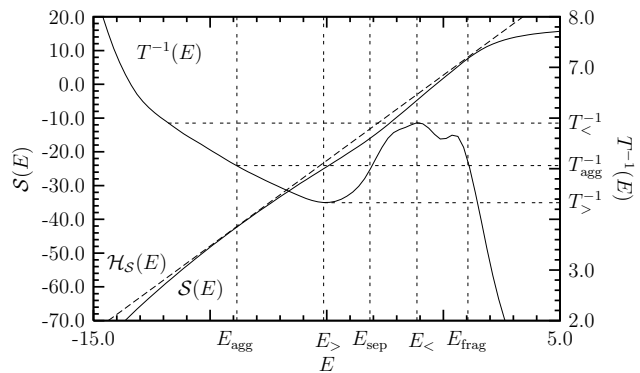}}
\caption{\label{fig:entropy} Microcanonical Hertz entropy ${\cal S}(E)$ of the 2$\times$S1 system,
concave Gibbs hull ${\cal H}_{\cal S}(E)$, and inverse caloric temperature $T^{-1}(E)$ as functions of 
energy. The phase separation regime ranges from $E_{\rm agg}$ to $E_{\rm frag}$. 
Between $T^{-1}_<$ and $T^{-1}_>$, the temperature is no suitable external control parameter and
the canonical interpretation is not useful: The inverse caloric temperature $T^{-1}(E)$ exhibits an
obvious backbending in the transition region. Note the second, less-pronounced backbending
in the energy range $E_<<E<E_{\rm frag}$.}
\end{figure}

In the microcanonical analysis, the system energy $E$ is kept (almost) fixed and treated as
an external control parameter. The system can only take macrostates with energies in the interval
$(E,E+\Delta E)$ with $\Delta E$ being sufficiently small to satisfy $\Delta G(E)=g(E)\Delta E$,
where $\Delta G(E)$ is the phase space volume of this energetic shell. In the limit $\Delta E\to 0$,
the total phase space volume up to the energy $E$ can thus be expressed as 
\begin{equation}
\label{eq:psvol}
G(E)=\int_{E_{\rm min}}^EdE'\,g(E'). 
\end{equation}
Since $g(E)$ is positive for all $E$, $G(E)$ is a monotonically 
increasing function and this quantity is suitably related to the microcanonical 
entropy ${\cal S}(E)$ of the system. In the definition of Hertz, 
\begin{equation}
\label{eq:hertz}
{\cal S}(E)=k_B\ln\,G(E).
\end{equation}
Alternatively, the entropy is often directly related to the density of states $g(E)$ and defined
as 
\begin{equation}
\label{eq:ent}
S(E)=k_B\ln\,g(E). 
\end{equation}
The density of states exhibits a decrease much faster than exponential towards the low-energy 
states. For this reason, the phase-space volume at
energy $E$ is strongly dominated by the number of states in the energy shell $\Delta E$. Thus 
$G(E)\approx \Delta G(E)\sim g(E)$ is directly related to the density of states. This virtual
identity breaks down in the higher-energy region, where $\ln\, g(E)$ is getting flat~-- in our case far
above the energetic regions being relevant for the discussion of the aggregation transition
(i.e., for energies $E\gg E_{\rm frag}$, see Fig.~\ref{fig:entropy}).
Actually, both definitions of the entropy led in our study to virtually
identical results in the analysis of the aggregation transition~\cite{jbj1}.
The (reciprocal) slope of the microcanonical entropy fixes the temperature scale and the corresponding 
caloric temperature is then defined via 
$T(E)=(\partial {\cal S}(E)/\partial E)^{-1}$ for fixed volume $V$ and particle number $N_{\rm tot}$.

As long as the mapping between the caloric temperature $T$ and the system energy $E$ is bijective,
the canonical analysis of crossover and phase transitions is suitable since the temperature can be
treated as external control parameter. For systems, where this condition is not satisfied, however, 
in a standard canonical analysis one may easily miss a physical effect accompanying 
condensation processes: Due to surface effects (the formation of the contact surface between the peptides
requires a rearrangement of monomers in the surfaces of the individual peptides), 
additional energy does not necessarily lead to
an increase of temperature of the condensate. Actually, the aggregate can even become colder. 
The supply of additional energy supports the fragmentation of parts of the aggregate, but this is 
overcompensated by cooperative processes of the particles aiming to reduce the surface tension.
Condensation processes are phase-separation processes and as such aggregated and fragmented phases
coexist. Since in this phase-separation region $T$ and $E$ are not bijective, this phenomenon is
called the ``backbending effect''. The probably most important class of systems exhibiting this 
effect is characterized by their smallness and the capability to form aggregates, depending on the 
interaction range. The fact that this effect could be indirectly observed in
sodium clustering experiments~\cite{schmidt1} gives rise to the hope
that backbending could also be observed in aggregation processes of small peptides.

Since the 2$\times$S1 system apparently belongs to this class, the backbending effect is
also observed in the aggregation/fragmentation transition of this system. This is shown 
in Fig.~\ref{fig:entropy}, where the microcanonical entropy ${\cal S}(E)$ is plotted as function 
of the system energy. The phase-separation region of aggregated and fragmented conformations
lies between $E_{\rm agg}\approx -8.85$ and $E_{\rm frag}\approx 1.05$.
Constructing the concave Gibbs
hull ${\cal H}_{\cal S}(E)$ by linearly connecting ${\cal S}(E_{\rm agg})$ and ${\cal S}(E_{\rm frag})$ 
(straight dashed line in Fig.~\ref{fig:entropy}), the entropic deviation due to surface effects is
simply $\Delta {\cal S}(E)={\cal H}_{\cal S}(E)-{\cal S}(E)$. The deviation is maximal for  
$E=E_{\rm sep}$ and $\Delta {\cal S}(E_{\rm sep})\equiv \Delta {\cal S}_{\rm surf}$ is the
surface entropy. The Gibbs hull also defines the aggregation transition temperature
\begin{equation}
\label{eq:tagg}
T_{\rm agg}=(\partial {\cal H}_{\cal S}(E)/\partial E)^{-1}.
\end{equation}
For the 2$\times$S1 system, we find $T_{\rm agg}\approx 0.198$, which is virtually
identical with the peak temperatures of the fluctuating quantities discussed in Sect.~\ref{sec:can}. 

The inverse caloric temperature $T^{-1}(E)$ is also plotted into Fig.~\ref{fig:entropy}. For a fixed
temperature in the interval $T_< <T< T_>$ ($T_< \approx 0.169$ and $T_> \approx 0.231$), 
different energetic macrostates coexist. This is a consequence
of the backbending effect. Within the backbending region, the temperature decreases with increasing
system energy. The horizontal line at $T^{-1}_{\rm agg}\approx 5.04$ is the Maxwell construction, i.e.,
the slope of the Gibbs hull ${\cal H}_{\cal S}(E)$. Although the transition seems to have similarities
with the van der Waals description of the condensation/evaporation transition of gases~-- the ``overheating''
of the aggregate between $T_{\rm agg}$ and $T_>$ (within the energy interval 
$E_{\rm agg}<E<E_{\rm >}\approx -5.13$) 
is as apparent as the ``undercooling'' of the fragments between $T_<$ and $T_{\rm agg}$ 
(in the energy interval $E_{\rm frag}>E>E_{\rm <}\approx -1.13$)~-- it is important to notice that in contrast to
the van der Waals picture the backbending effect in-between is a real physical effect. Another
essential result is that in the transition region the temperature is not a suitable external control 
parameter: The macrostate of the system cannot be adjusted by fixing the temperature. The better choice
is the system energy which is unfortunately difficult to control in experiments. Another direct 
consequence of the energetic ambiguity for a fixed temperature between $T_<$ and $T_>$ is that the
canonical interpretation is not suitable for detecting the backbending phenomenon. 
It should also be noted that in this region, the microcanonical
specific heat $c_V(E)$ can become negative~\cite{jbj1}, which is a remarkable, but somehow 
``exotic'' side effect.

\begin{figure}
\centerline{\epsfxsize=8.8cm \epsfbox{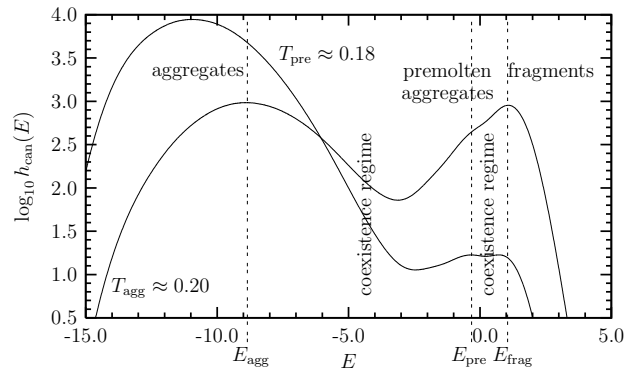}}
\caption{\label{fig:canhist} 
Logarithmic plots of the canonical energy histograms (not normalized) at $T\approx 0.18$ and 
$T\approx 0.20$, respectively.}
\end{figure}

The precise microcanonical analysis reveals also a further detail of the aggregation transition. Close to
$E_{\rm pre}\approx -0.32$, the $T^{-1}$ curve in Fig.~\ref{fig:entropy} exhibits another
``backbending'' which signalizes a second, but unstable transition of the same type. 
The associated transition temperature $T_{\rm pre}\approx 0.18$ is smaller than $T_{\rm agg}$,
but this transition occurs in the energetic region where fragmented states dominate. Thus
this transition can be interpreted as the premelting of aggregates by forming intermediate states. 
These intermediate structures are rather weakly stable: The population of the premolten aggregates
never dominates. In particular, at $T_{\rm pre}$, where premolten aggregates and 
fragments coexist, the population of compact aggregates is much larger. This can nicely be seen in
the canonical energy histograms at these temperatures plotted in Fig.~\ref{fig:canhist}, where
the second backbending is only signalized by a small cusp in the coexistence region. Since both 
transitions are phase-separation processes, structure formation is accompanied by releasing latent
heat which can be defined as the energetic widths of the phase coexistence regimes, i.e., 
$\Delta Q_{\rm agg}=E_{\rm frag}-E_{\rm agg}=T_{\rm agg}[{\cal S}(E_{\rm frag})-{\cal S}(E_{\rm agg})]\approx 9.90$ 
and $\Delta Q_{\rm pre}=E_{\rm frag}-E_{\rm pre}=T_{\rm pre}[{\cal S}(E_{\rm frag})-{\cal S}(E_{\rm pre})]\approx 1.37$.
Obviously, the energy required to melt the premolten aggregate is much smaller than to dissolve a 
compact (solid) aggregate. 

For the comparison of the surface entropies, we use the definition~(\ref{eq:ent}) of the entropy.
In the case of the aggregation transition, the surface entropy is 
$\Delta {\cal S}_{\rm surf}^{\rm agg} \approx \Delta S_{\rm surf}^{\rm agg}= H_S(E_{\rm sep})-S(E_{\rm sep})$,
where $H_S(E)\approx{\cal H}_{\cal S}(E)$ is the concave Gibbs hull of $S(E)$.
Since $H_S(E_{\rm sep})=H_S(E_{\rm frag})-(E_{\rm frag}-E_{\rm sep})/T_{\rm agg}$ and 
$H_S(E_{\rm frag})=S(E_{\rm frag})$, the surface entropy is
\begin{equation}
\label{eq:surfent}
\Delta S_{\rm surf}^{\rm agg}=S(E_{\rm frag})-S(E_{\rm sep}) - \frac{1}{T_{\rm agg}}(E_{\rm frag}-E_{\rm sep}).
\end{equation}
Yet utilizing that the canonical distribution $h_{\rm can}(E)$ at $T_{\rm agg}$ shown 
in Fig.~\ref{fig:canhist} is $h_{\rm can}(E)\sim g(E)\exp(-E/k_BT_{\rm agg})$,
the surface entropy can be written in the simple and computationally convenient form~\cite{wj1}:
\begin{equation}
\label{eq:surfent2}
\Delta S_{\rm surf}^{\rm agg}=k_B\ln\,\frac{h_{\rm can}(E_{\rm frag})}{h_{\rm can}(E_{\rm sep})}.
\end{equation}
A similar expression is valid for the coexistence of premolten and fragmented states at $T_{\rm pre}$
The corresponding canonical distribution is also shown in Fig.~\ref{fig:canhist}. Thus, we obtain
(in units of $k_B$) for the surface entropy of the aggregation transition 
$\Delta S_{\rm surf}^{\rm agg}\approx 2.48$ and for the premelting $\Delta S_{\rm surf}^{\rm pre}\approx 0.04$,
confirming the weakness of the interface between premolten aggregates and fragmented states.
\section{\label{sec:larger} Aggregation transition in larger heteropolymer systems}
In order to verify the general validity of the statements in the previous section for the 
2$\times$S1 system, we have also performed simulations of systems consisting of three (in the
following referred to as 3$\times$S1) and four (4$\times$S1) identical peptides with sequence S1.

\begin{figure}
\centerline{\epsfxsize=8.8cm \epsfbox{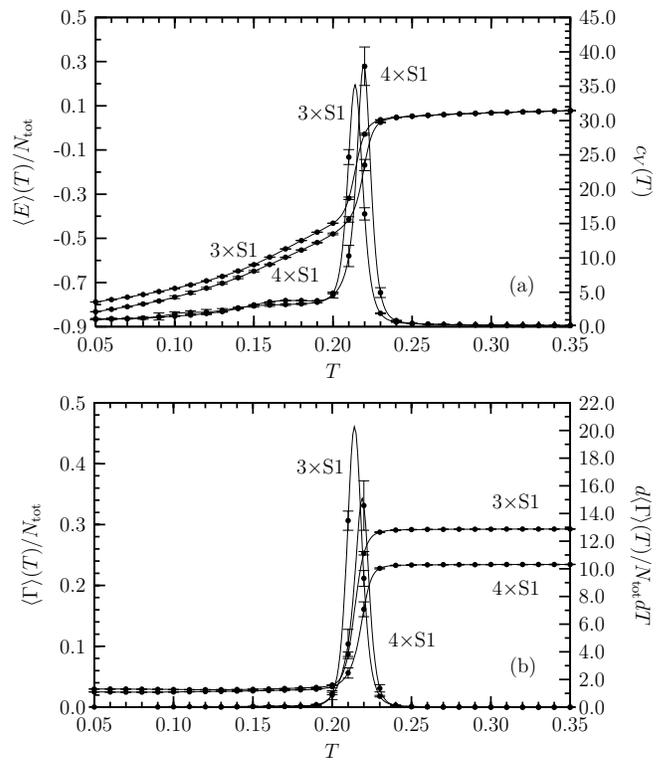}}
\caption{\label{fig:canon3_4x13.1} (a) Mean energy $\langle E\rangle/N_{\rm tot}$ and specific heat
per monomer $c_V$, (b) mean aggregation parameter $\langle \Gamma\rangle/N_{\rm tot}$ and its fluctuations 
$d\langle\Gamma\rangle/N_{\rm tot}dT$ as functions
of the temperature for the 3$\times$S1 and 4$\times$S1 heteropolymer systems.}
\end{figure}

Although the formation of compact hydrophobic cores is more complex in larger compounds
of our exemplified sequence S1, the aggregation transition is little influenced by this.
This is nicely seen in Figs.~\ref{fig:canon3_4x13.1}(a) and~\ref{fig:canon3_4x13.1}(b), where
the temperature dependence of the canonical expectation values of $\Gamma$ and $E$, 
as well as for their fluctuations, are shown for the 3$\times$S1 system. For comparison, also 
results for the 4$\times$S1 system are plotted into the same figures.
Note that for the 4$\times$S1 system finite-size effects are
larger since, for computational reasons, we have kept the edge length of the simulation box $L=40$,
which is smaller than the successive arrangement of four straight chains with 13 monomers.
This influences primarily the entropy in the high-energy regime far above the aggregation 
transition energy. Nonetheless, in the canonical interpretation, it acts back onto the transition
as undesired states (chain ends overlapping due to the periodic boundary conditions)
are (weakly) populated at the transition temperature, whereas others are suppressed. We have performed a 
detailed analysis of the box size dependence (results not shown) and found that the canonical
transition temperature scales slightly, but noticeably with the box size. 
Thus, the results obtained by
\emph{canonical} statistics for the 4$\times$S1 system should not quantitatively be compared to 
the canonical results for the 2$\times$S1 and 3$\times$S1 systems. 
 
As has already been discussed for the 2$\times$S1 system, there are also for the larger systems
no obvious signals for separate
aggregation and hydrophobic-core formation processes. Only weak activity in the energy
fluctuations in the temperature region below the aggregation transition temperature
indicates that local restructuring processes of little cooperativity 
(comparable with the discussion of the premolten aggregates in the discussion of the
2$\times$S1 system) are still happening. The strength of the aggregation transition is
also documented by the fact that the peak temperatures of energetic \emph{and} aggregation parameter
fluctuations are virtually identical for the 3$\times$S1 system, i.e., the aggregation 
temperature is $T_{\rm agg}\approx 0.21$ (for 4$\times$S1 $T_{\rm agg}\approx 0.22$).

For homogeneous multiple-chain systems two variants of thermodynamic limits are of
particular interest: (i) $M\to\infty$, while $N_\mu={\rm const}$, 
(ii) $N_\mu\to\infty$ with $M={\rm const}$; both limits considered for constant polymer density. 
Since for proteins the sequence of
amino acids is fixed, in this case only (i) is relevant and it is future work to
perform a scaling analysis for multiple-peptide systems in this limit. A particularly
interesting question is to what extent remnants of the finite-system effects, as discussed in this
paper, survive in the limit of an infinite number of chains, dependent on the peptide density.
Since we have focused our study on the precise analysis of systems of few peptides for all energies
and temperatures, it was computationally inevitable to restrict ourselves to small systems, for which
a scaling analysis is not very useful. Nonetheless, we would like to devote a few interesting 
remarks to the comparison of, once more, microcanonical aspects of the aggregation transition in 
dependence of the system size.

\begin{figure}
\centerline{\epsfxsize=8.8cm \epsfbox{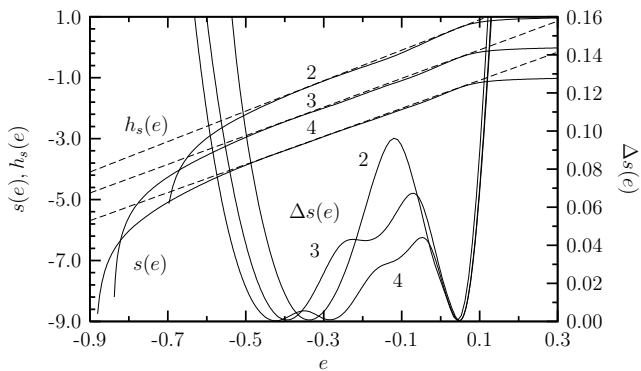}}
\caption{\label{fig:ent234} 
Microcanonical entropies per monomer $s(e)$, respective Gibbs constructions $h_s(e)$ (left scale), and 
deviations $\Delta s(e)=h_s(e)-s(e)$ (right scale) for 2$\times$S1 (labeled as 2), 3$\times$S1 (3), and 
4$\times$S1 (4) as functions of the energy per monomer $e$.}
\end{figure}
In Fig.~\ref{fig:ent234}, the microcanonical entropies per monomer $s(e)={\cal S}(e)/N_{\rm tot}$ 
(shifted by an unimportant constant for clearer visibility) and the corresponding
Gibbs hulls $h_s(e) = {\cal H}_{\cal S}(e)/N_{\rm tot}$ are shown for 2$\times$S1 (in the figure
denoted by ``2''), 3$\times$S1 (``3''), and 4$\times$S1 (``4''),
respectively, as  functions of the energy per monomer $e=E/N_{\rm tot}$. 
Although the convex entropic ``intruder'' is apparent for larger
systems as well, its relative strength decreases with increasing number of chains. The slopes
of the respective Gibbs constructions determine the aggregation temperature~(\ref{eq:tagg}) 
which are found to be $T_{\rm agg}^{3\times{\rm S}1}\approx 0.212$ and 
$T_{\rm agg}^{4\times{\rm S}1}\approx 0.217$ confirming the peak temperatures of the fluctuation
quantities plotted in Fig.~\ref{fig:canon3_4x13.1}. 

The existence of the interfacial boundary entails a transition barrier whose strength is characterized
by the surface entropy $\Delta {\cal S}_{\rm surf}$. In Fig.~\ref{fig:ent234},  the 
individual entropic deviations per monomer, $\Delta s(e)=\Delta {\cal S}(e)/N_{\rm tot}$ 
are also shown and the maximum deviations, i.e., 
the surface entropies $\Delta {\cal S}_{\rm surf}$ and relative surface entropies per monomer
$\Delta s_{\rm surf}=\Delta {\cal S}_{\rm surf}/N_{\rm tot}$
are listed in Table~{\ref{tab:lsys}}.
There is no apparent difference between the values of $\Delta {\cal S}_{\rm surf}$ 
that would indicate a trend for a vanishing 
of the \emph{absolute} surface barrier in larger systems. However, the \emph{relative} surface entropy
$\Delta s_{\rm surf}$
obviously decreases. Whether or not it vanishes in the thermodynamic limit cannot be decided
from our results and is a study worth in its own right. 

\begin{table}
\caption{\label{tab:lsys} Aggregation temperatures $T_{\rm agg}$, surface entropies 
$\Delta {\cal S}_{\rm surf}$, relative surface entropies per monomer $\Delta s_{\rm surf}$,
relative aggregation and fragmentation energies per monomer, $e_{\rm agg}$ and $e_{\rm frag}$,
respectively, latent heat per monomer $\Delta q$, and phase-separation entropy per monomer
$\Delta q/T_{\rm agg}$. All quantities for systems consisting of two, three, and four 
13mers with AB sequence S1.}
\begin{ruledtabular}
\begin{tabular}{c|ccccccc}
system & $T_{\rm agg}$ & $\Delta {\cal S}_{\rm surf}$ & $\Delta s_{\rm surf}$ & 
$e_{\rm agg}$ & $e_{\rm frag}$ & $\Delta q$ & $\Delta q/T_{\rm agg}$\\ \hline
2$\times$S1 & 0.198 & 2.48 & 0.10 & $-0.34$ & 0.04 & 0.38 & 1.92 \\
3$\times$S1 & 0.212 & 2.60 & 0.07 & $-0.40$ & 0.05 & 0.45 & 2.12 \\
4$\times$S1 & 0.217 & 2.30 & 0.04 & $-0.43$ & 0.05 & 0.48 & 2.21 \\
\end{tabular}
\end{ruledtabular}
\end{table}
It is also interesting that subleading effects increase 
and the double-well form found for 2$\times$S1 changes by higher-order effects,
and it seems that for larger systems the almost single-step aggregation of 2$\times$S1
is replaced by a multiple-step process.

Not surprisingly,
the fragmented phase is hardly influenced by side effects and the rightmost minimum 
in Fig.~\ref{fig:ent234} lies well at
$e_{\rm frag}= E_{\rm frag}/N_{\rm tot}\approx 0.04-0.05$. Since the 
Gibbs construction 
covers the whole convex region of $s(e)$, the aggregation energy per monomer
$e_{\rm agg}= E_{\rm agg}/N_{\rm tot}$ 
corresponds to the leftmost minimum and its value changes noticeably with the number of chains.
In consequence, the latent heat per monomer 
$\Delta q = \Delta Q/N_{\rm tot}=T_{\rm agg}[{\cal S}(E_{\rm frag})-{\cal S}(E_{\rm agg})]/N_{\rm tot}$ that 
is required to fragment the aggregate increases from two to four chains in the 
system (see Table~\ref{tab:lsys}). 
Although the systems under consideration are 
too small to extrapolate phase transition properties in the thermodynamic limit, 
it is obvious that the aggregation-fragmentation transition
exhibits strong similarities to condensation-evaporation transitions of colloidal systems. Given that,
the entropic transition barrier $\Delta q/T_{\rm agg}$, which we see increasing with the number of
chains (cf.\ the values in Table~\ref{tab:lsys}), would survive in the thermodynamic limit 
and the transition was 
first-order-like. More surprising would be, however, if the convex intruder would not 
disappear, i.e., if the absolute and relative surface entropies 
$\Delta {\cal S}_{\rm surf}$ and $\Delta s_{\rm surf}$ do not vanish. This is definitely
a question of fundamental interest as the common claim is that pure surface effects typically
exhibited only by ``small'' systems are irrelevant in the thermodynamic limit. This requires, however,
studies of much larger systems. It should clearly be noted, however, that protein aggregates forming
themselves in biological systems often consist only of a few peptides and are definitely of small 
size and the surface effects are responsible for structure formation and are not unimportant
side effects. One should keep in mind that standard thermodynamics and the thermodynamic 
limit are somewhat theoretical constructs valid only for very large systems. 
The increasing interest in physical properties of small systems, in particular in conformational transitions
in molecular systems, requires in part a revision of dogmatic thermodynamic views. Indeed, by 
means of today's chemo-analytical and experimental equipment, effects like those described
throughout the paper, should actually experimentally be verifiable as these are real physical
effects. For studies of the condensation of atoms, where a similar behavior occurs, 
such experiments have actually already been performed~\cite{schmidt1}.
\section{\label{sec:sum}Summary}
In this paper, we have extended the microcanonical analysis of the aggregation of an
exemplified two-peptide system~\cite{jbj1} by interpreting the results from the 
multicanonical and the canonical perspective as well. In addition, these results are compared with 
aggregation properties of larger systems consisting of three and four peptides,
each of which with the same sequence. From the conventional canonical analysis of statistical
fluctuations of energy and a suitably chosen ``order'' parameter -- the root mean square distance
of the centers of masses of the individual polymers -- we obtain the typical small-system 
indications of a thermodynamic phase transition: Sharp peaks in the specific heat and in the
order parameter fluctuations at almost the same temperature signalize a strong transition,
which we clearly identify as the aggregation transition. For all systems considered, the
general behavior is similar. There is only this single transition which also indicates that 
conformational changes of the polymers accompany the aggregation process and are not separate 
transitions. We expect that this coincidence is sequence-dependent and a comparison between
different sequences is a study in its own right. At least for the semiflexible homopolymer of
same size which in our notation would have the sequence $A_{13}$ (or also $B_{13}$), we find 
that aggregation and collapse are separate processes~\cite{jbj2}. 

A quite remarkable result of the exemplified heteropolymer study presented in this paper is 
that the aggregation process of a small number of peptides is a phase-separation process, where
interfacial surface effects entail a loss of entropy. This loss must be compensated by 
additional energy delivered to the system. In consequence, the caloric temperature decreases, i.e.,
the aggregate is getting colder although its total energy increases. This is known as the 
temperature ``backbending'' which is a real thermodynamic effect and not an artefact of the theory. 
In the systems considered throughout the paper, the relative influence of the surface effects
seems to decrease with the number of chains in the system. Since the length of the peptides is fixed
by their hydrophobic-polar monomer composition, a thermodynamic limit towards infinite chain lengths
is, however, not existing. It is just the smallness of such molecular systems that allow these
to trigger biological interchange processes which are inevitably connected with conformational activity.
\acknowledgments
This work is partially supported by the DFG (German Science Foundation) under Grant
No.\ JA 483/24-1/2. M.B.\ thanks the DFG and the Wenner-Gren Foundation for research fellowships.
Support by the DAAD-STINT Personnel Exchange Programme with Sweden is gratefully
acknowledged.
We also thank the John von Neumann Institute for Computing (NIC), 
Forschungszentrum J{\"u}lich, for a supercomputer time grant under No.~hlz11. 

\begin{thebibliography}{99}
%
\bibitem{gsponer1}
J.\ Gsponer and M.\ Vendruscolo, Prot.\ \& Pept.\ Lett.\ \textbf{13}, 287 (2006).
%
\bibitem{lin1}
H.\ Lin, R.\ Bhatia, and R.\ Lal, FASEB J.\ \textbf{15}, 2433 (2001).
%
\bibitem{quist1}
A.\ Quist, I.\ Doudevski, H.\ Lin, R.\ Azimova, D.\ Ng, B.\ Frangione, B.\ Kagan, J.\ Ghiso,
and R.\ Lal, Proc.\ Natl.\ Acad.\ Sci.\ (USA) \textbf{102}, 10427 (2005).
%
\bibitem{lashuel1}
H.\ A.\ Lashuel and P.\ T.\ Lansbury Jr., Quart.\ Rev.\ Biophys.\ \textbf{39}, 167 (2006).
%
\bibitem{gross1}
D.\ H.\ E.\ Gross, \emph{Microcanonical Thermodynamics} (World Scientific, Singapore, 2001).
%
\bibitem{gross2}
D.\ H.\ E.\ Gross and J.\ F.\ Kenney, J.\ Chem.\ Phys.\ \textbf{122}, 224111 (2005).
%
\bibitem{thirring1}
W.\ Thirring, Z.\ Physik \textbf{235}, 339 (1970).
%
\bibitem{schmidt1}
M.\ Schmidt, R.\ Kusche, T.\ Hippler, J.\ Donges, W.\ Kronm\"uller, B.\ von Issendorff, and
H.\ Haberland, Phys.\ Rev.\ Lett.\ \textbf{86}, 1191 (2001). 
%
\bibitem{pichon1}
M.\ Pichon, B.\ Tamain, R.\ Bougault, and O.\ Lopez, Nucl.\ Phys.\ A \textbf{749}, 93c (2005).
%
\bibitem{lopez1}
O.\ Lopez, D.\ Lacroix, and E.\ Vient, Phys.\ Rev.\ Lett.\ \textbf{95}, 242701 (2005).
%
\bibitem{wj1}
W.\ Janke, Nucl.\ Phys.\ B (Proc.\ Suppl.) \textbf{63A-C}, 631 (1998).
%
\bibitem{pleimling1}
H.\ Behringer and M.\ Pleimling, Phys.\ Rev.\ E \textbf{74}, 011108 (2006).
%
\bibitem{wales1}
D.\ J.\ Wales and R.\ S.\ Berry, Phys.\ Rev.\ Lett.\ \textbf{73}, 2875 (1994);
D.\ J.\ Wales and J.\ P.\ K.\ Doye, J.\ Chem.\ Phys.\ \textbf{103}, 3061 (1995).
%
\bibitem{hilbert1}
S.\ Hilbert and J.\ Dunkel, Phys.\ Rev.\ E \textbf{74}, 011120 (2006);
J.\ Dunkel and S.\ Hilbert, Physica A \textbf{370}, 390 (2006).
%
\bibitem{nussbaumer1}
A.\ Nu{\ss}baumer, E.\ Bittner, T.\ Neuhaus, and W.\ Janke, Europhys.\ Lett.\ \textbf{75},
716 (2006).
%
\bibitem{gross3}
D.\ H.\ E.\ Gross, Physica E \textbf{29}, 251 (2005).
%
\bibitem{jbj1}
C.\ Junghans, M.\ Bachmann, and W. Janke, Phys.\ Rev.\ Lett.\ \textbf{97}, 218103 (2006).
%
\bibitem{still1}
F.\ H.\ Stillinger, T.\ Head-Gordon, and C.\ L.\ Hirshfeld, Phys.\ Rev.\ E \textbf{48}, 1469 (1993);
F.\ H.\ Stillinger and T.\ Head-Gordon, Phys.\ Rev.\ E \textbf{52}, 2872 (1995).
%
\bibitem{ssbj1}
S.\ Schnabel, M.\ Bachmann, and W.\ Janke, Phys.\ Rev.\ Lett.\ \textbf{98}, 048103 (2007);
J.\ Chem.\ Phys.\ \textbf{126}, 105102 (2007).
%
\bibitem{baj1}
M.\ Bachmann, H.\ Ark{\i}n, and W.\ Janke, Phys.\ Rev.\ E \textbf{71}, 031906 (2005).
%
\bibitem{muca1}
B.\ A.\ Berg and T.\ Neuhaus, Phys.\ Lett.\ B \textbf{267}, 249 (1991);
Phys.\ Rev.\ Lett.\ \textbf{68}, 9 (1992).
%
\bibitem{muca2}
W.\ Janke, Physica A \textbf{254}, 164 (1998); B.\ A.\ Berg, Fields Inst.\ Comm. \textbf{26}, 1 (2000).
%
\bibitem{muca3}
W.\ Janke, {\em Histograms and All That},
in: {\em Computer Simulations of Surfaces and Interfaces}, NATO Science Series, II.\
Mathematics, Physics and Chemistry -- Vol.\ \textbf{114},
edited by B.\ D\"unweg, D.\ P.\ Landau, and A.\ I.\ Milchev (Kluwer, Dordrecht, 2003), p.\ 137.
%
\bibitem{celik1}
T.\ \c{C}elik and B.\ A.\ Berg, Phys.\ Rev.\ Lett.\ \textbf{69}, 2292 (1992).
%
\bibitem{pt1}
K.\ Hukushima and K.\ Nemoto, J.\ Phys.\ Soc.\ Jap.\ \textbf{65}, 1604 (1996);
K.\ Hukushima, H.\ Takayama, and K.\ Nemoto, Int.\ J.\ Mod.\ Phys.\ C \textbf{7}, 337 (1996).
%
\bibitem{pt2}
C.\ J.\ Geyer, in: \emph{Computing Science and Statistics}, Proceedings of the 23rd
Symposium on the Interface, edited by E.\ M.\ Keramidas (Interface Foundation,
Fairfax Station, 1991), p.\ 156.
%
%
\bibitem{elp}
U.\ H.\ E.\ Hansmann and L.\ T.\ Wille, Phys.\ Rev.\ Lett.\ \textbf{88}, 068105 (2002).
%
\bibitem{bj1}
M.\ Bachmann and W.\ Janke, Phys.\ Rev.\ Lett.\ \textbf{91}, 208105 (2003);
J.\ Chem.\ Phys.\ \textbf{120}, 6779 (2004).
%
\bibitem{jbj2}
C.\ Junghans, M.\ Bachmann, and W. Janke, unpublished.
%


\end{thebibliography}
\end{document}